%% file: ieee.tex
\documentclass[conference]{IEEEtran}
\IEEEoverridecommandlockouts
\usepackage{cite}
\usepackage{amsmath,amssymb,amsfonts}
\usepackage{algorithmic}
\usepackage{graphicx}
\usepackage{textcomp}
\usepackage{xcolor}
\usepackage{tabu}
\usepackage{subfigure}
\usepackage{balance}

\def\BibTeX{{\rm B\kern-.05em{\sc i\kern-.025em b}\kern-.08em
    T\kern-.1667em\lower.7ex\hbox{E}\kern-.125emX}}
\usepackage[letterpaper,%
            left=1.05in,right=1.02in,top=0.75in,bottom=1in,%
            footskip=.25in]{geometry}
\begin{document}

\title{Context Learning for Bone Shadow Exclusion in CheXNet Accuracy Improvement \\}
%Radiological Diagnosis of Lung Disease\\}

\author{\IEEEauthorblockN{Minh-Chuong Huynh}
\IEEEauthorblockA{\textit{Faculty of Information Technology} \\
\textit{University of Science, VNU-HCM}\\
Ho Chi Minh, Vietnam \\
1412060@student.hcmus.edu.vn}
\and
\IEEEauthorblockN{Trung-Hieu Nguyen}
\IEEEauthorblockA{\textit{Faculty of Information Technology} \\
\textit{University of Science, VNU-HCM}\\
Ho Chi Minh, Vietnam \\
1412165@student.hcmus.edu.vn}
\and
\IEEEauthorblockN{Minh-Triet Tran}
\IEEEauthorblockA{\textit{Faculty of Information Technology} \\
\textit{University of Science, VNU-HCM}\\
Ho Chi Minh, Vietnam \\
tmtriet@fit.hcmus.edu.vn}
}

\maketitle

\begin{abstract}
\input{abstract}
\end{abstract}

\begin{IEEEkeywords}
medical image processing, bone shadow exclusion, CheXNet, X-ray images, radiological diagnosis
\end{IEEEkeywords}

\section{Introduction}
\input{introduction}

\section{Background and Related Work}
\input{previous_work}

\section{Proposed Method}
\input{bxf}

\section{Experiments}
\input{experiment}

\section{Conclusion}
\input{conclusion}

\section*{Acknowledgment}
\input{ack}

\bibliographystyle{IEEEtran}
\bibliography{refs}

\end{document}

%% file: abstract.tex
Chest X-ray examination plays an important role in lung disease detection. The more accuracy of this task, the more experienced radiologists are required. After ChestX-ray14 dataset containing over 100,000 frontal-view X-ray images of 14 diseases was released, several models were proposed with high accuracy. In this paper, we develop a work flow for lung disease diagnosis in chest X-ray images, which can improve the average AUROC of the state-of-the-art model from 0.8414 to 0.8445. We apply image preprocessing steps before feeding to the 14 diseases detection model. Our project includes three models: the first one is DenseNet-121 to predict whether a processed image has a better result, a convolutional auto-encoder model for bone shadow exclusion is the second one, and the last is the original CheXNet. 

%% file: introduction.tex
Lung cancer is the leading cause of death worldwide, accounting for 1.69 million cases in 2015 \cite{who:cancerfact}, more than breast, colon and prostate cancers combined. Funding for lung cancer research is critical due to the illness's prominence and this disease is often diagnosed in later stages, when it is less treatable. New advances, including deep neural networks, hold great promise for screening, early detection and personalized therapies. Furthermore, lung cancer does not have to be fatal, groundbreaking new treatments dramatically alter lung cancer survival rates every day \cite{lung:cancerfact}. Previous lung diseases can lead to lung cancer in the future. The history of COPD, chronic bronchitis or emphysema, and pneumonia are top diseases which have a high probability of causing lung cancer \cite{brenner2011previous}. 

Pneumonia is a serious lung infection, and is also the second most common cause of death in people with lung cancer. This disease can be mild and in most cases only require a few weeks of treatment. A small number of cases require several weeks of staying in the hospital or this disease can also be life-threatening and even fatal. Higher than cigarette smoking, lung cancer patients get a high risk of developing pneumonia. Symptoms of pneumonia in lung cancer cases are tough to identify with normal medical checks, such as physical exam, listening to patients' chest with a stethoscope while breathing \cite{lung:pneumonia}. Usually, doctors require imaging studies, and chest X-ray is the first priority choice. Looking at lung structure and chest cavity in X-ray images is an effective way for medical experts to make decisions. 

CheXNet \cite{rajpurkar2017chexnet} was published on 14th November 2017 by Standford Machine Learning group. This algorithm is the state-of-the-art method for radiological diagnosis of lung disease, especially since it has a higher accuracy in Pneumonia compared with experienced radiologists. Using a very deep neural network, it improves the accuracy of previous methods to 0.8414. However, we realize that preprocessing the training dataset is an essential work that is not applied to the current model. 

X-ray image enhancement and bone suppression help improve the accuracy of diagnosing. In most cases, lung cancer lesions that are missed on frontal-view chest radiological images are situated behind ribs, and soft tissue images can improve the performance of medical experts in disease detection \cite{gusarev2017deep}. Moreover, the contrast of X-ray images is usually low \cite{huang2016noise}, degrading the image quality as well as decreasing accuracy when feeding directly to CheXNet model. To remove these noises, we build a process for improving CheXNet accuracy by preprocessing input images.

\begin{figure*}[!htp]
\centering
	\vspace{0.05in}
	\includegraphics[width=\textwidth]{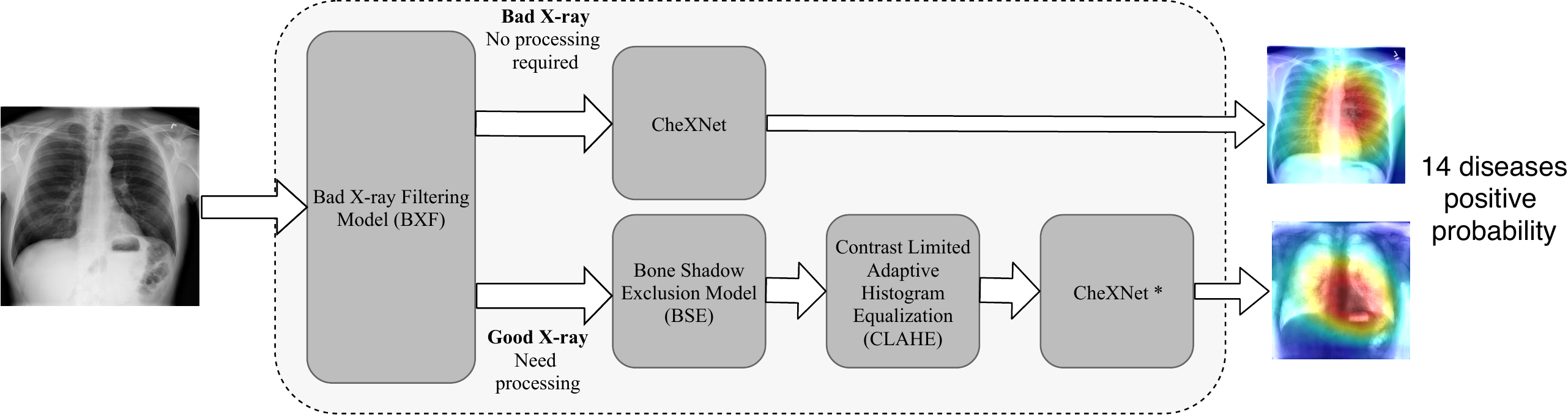}
	\label{fig:workflow}
    \caption{Our flow includes a filtering model which predict whether an input image have a high accuracy when preprocessing. The (preprocessed) image is fed to CheXNet model for diagnosing 14 lung diseases. The CheXNet* is trained with bone shadow exclusion version of ChestX-ray14 dataset}. 
\end{figure*}

Our flow (shown in Figure \ref{fig:workflow}) defines the end-to-end process that inputs a chest X-ray image and outputs the probability of 14 lung diseases with a heatmap localizing the areas of the (denoised) image most indicative of top injuries. The BXF model is built for binary classification with the idea from CheXNet, to decide whether the preprocessing of an image is essential. At the processing-denosing stage, a convolutional auto-encoder neural network is used for bone shadow exclusion from X-ray images. CheXNet is the main model for lung disease detection. We have trained two CheXNet models with two versions of ChestX-ray14 dataset: the published version and our preprocessed one.

The main contributions of our paper are as follows:
\begin{itemize}

\item Bone shadow exclusion and contrast enhancement are applied to X-ray images for clarifying the key features.
\item We propose a deep learning model to automatically decide whether preprocessing is necessary for each individually X-ray image.
\end{itemize}

The rest of the paper is organized as follows. In Section 2, we briefly review the background and related works. Our proposed workflow and context learning model for bone shadow exclusion are in Section 3. Experimental results on ChestX-ray14 dataset are discussed in Section 4. The final section is for conclusion and future work.

%% file: previous_work.tex
\subsection{Bone shadow exclusion}
Dual-energy subtraction (DES) \cite{vock2009dual} is an imaging technique to reduce bone shadow in chest X-ray images. DES involves capturing two radiographs with the use of two different energy-level X-ray exposures. Depending on the level, the output image highlights either soft-tissue or bone components; these images are combined to form the bone suppression. However, this DES system is not available in all hospitals. 

With deep learning development, we can use neural networks to construct soft-tissue images from normal chest X-rays. A three-layer convolutional neural network was used to construct the virtual soft tissue image \cite{yang2017cascade}. The authors have trained the model with 404 dual-energy exams to estimate and subtract the bone image from the input image. This method can also be applied to radiographs from different sources. Based on this idea, Gusarev et. al. \cite{gusarev2017deep} develops a an autoencoder-like convolutional model for bone shadow exclusion. By proposing a new loss function, this model uses fewer data but still keeps the high quality of output image.

The auto-encoder model is a stack of convolutional auto-encoders with encoder and decoder sharing the same but mirrored weights. The model includes three encoders, each encoding an image into 16, 32 and 64 neurons, respectively. The optimizer minimizes the mean squared error (MSE) along with maximizes multi-scale structural similarity (MS-SSIM) \cite{wang2003multi} of decoded image and a soft-tissue version (the ground truth). The final loss function is defined as this formula: 
\[L=\alpha.MSSI+(1-\alpha).MSE\] with \(alpha=0.84\) following Zhao et. al. \cite{zhao2017loss}

In previous research, the authors used 35 pairs of chest radiographs and its soft-tissue versions. Most of the soft tissue images are the results of dual-energy subtraction, and were acquired from different online sources. After getting 4000 image pairs from 35 initial one by using combined affine transformations, these images were cropped to 440x440 pixels before feeding to the auto-encoder model

\subsection{Contrast enhancement}
X-ray images usually have low contrast, degrading quality and affecting the diagnosis results of doctors. Historically, histogram equalization is a popular method, which distributes image intensity equally, increasing contrast in lower regions. However, this equalization works effectively if only histogram is not globally equally distributed, and can increase noise in image, blur or remove key features in X-ray images. 

Local histogram equalization can work well with X-ray images. Adaptive Histogram Equalization (AHE) is an example; it adjusts the contrast locally in small regions of image, increasing the contrast better than global equalization. However, this method still makes noise in output image. The Contrast Limited Adaptive Histogram Equalization (CLAHE), another version of AHE,  adjusts the contrast without putting more noise to the image. By defining a threshold when equalizing, the noise is reduced. Research from Huang et. al. \cite{huang2016noise}, they used "abolute mean brightness error" (AMB) to qualify the images after equalization between AHE and CLAHE. The result shows that AHE's error is 2.5 times higher than CLAHE's, 62.3860 and 23.6596 respectively. 

\subsection{CheXNet}
CheXNet is a 121-layer Dense Convolutional Neural Network (DenseNet) \cite{DenseNet2018} that is trained with ChestX-ray14. DenseNets help improve the flow of information and gradients through the network, increasing the number of parameters without overfitting like fully-connected networks. This kind of network works very well with X-ray images. The model is trained with the initial weights from ImageNet \cite{deng2009imagenet}, the unweighted binary cross entropy loss is used for optimization. CheXNet outperformed the best published results on all 14 pathologies in the ChestX-ray14 dataset. In detecting Mass, Nodule, Pneumonia, and Emphysema, CheXNet has a margin of \(>\) 0.05 AUROC over previous results. The authors also compared this deep learning model with radiologists having 4, 7, 25 and 28 years of experience. Surprisingly, CheXNet's performance is statistically higher than radiologists'. The state-of-the-art deep learning model can improve healthcare delivery and create opportunities for poor condition parts of the world to access to medical imaging expertise.

To interpret the network predictions, the authors also use class activation mappings (CAMs) \cite{zhou2016CAM} to visualize the areas of the image most indicative of the disease. These localizations in output images can help doctors deeply observe high-value areas.

%% file: bxf.tex
\subsection{Overview}
\begin{figure}
    \centering
    \vspace{0.05in}
    \includegraphics[width=0.45\columnwidth]{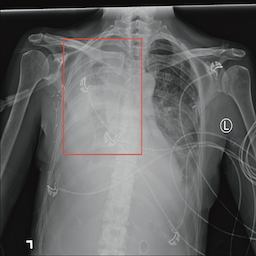}
        \includegraphics[width=0.45\columnwidth]{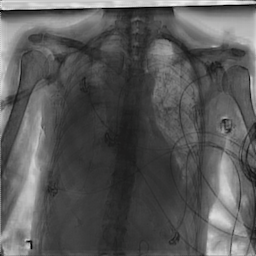}
    \caption{Images getting bad results when preprocessing}
    \label{analyse:bad}
\end{figure}
Figure \ref{fig:workflow} describes overall our flow in chest X-ray diagnosis. The Bad X-ray Filtering model (BXF) choose the appropriate strategy for each individual chest frontal-view radiograph, increasing the accuracy of CheXNet model on each single input. This decision is made completely automatic by previous learning key features from the dataset, no human being effort involves in this task.

Good images predicted by the BXF model will be diagnosed better, get higher accuracy with CheXNet model after preprocessing. In the other hand, the original version of bad images will be predicted better than the preprocessing version. Therefore, bad images will be fed directly to a CheXNet model. This model is trained with published ChestX-ray14 dataset. Besides that, another CheXNet model will receives good images which are chosen for preprocessing strategy. This model is trained with the bone suppression and contrast enhancement version of ChestX-ray14 (BSE version). There is no weights sharing between two models, each of them was trained with initial weights from a pretrained model of ImageNet.

Before observing the dataset carefully, we had expected that using only BSE version can get the better result. However, when suppressing bone shadow from normal X-Ray images, the training dataset misses some special types that currently exist on ChestX-ray14 (Figure \ref{analyse:bad}). Therefore, the BSE data cannot increase the quality of the CheXNet model individually.

\subsection{Context learning for bone shadow exclusion}
\begin{figure*}[!htb]
	\vspace{0.05in}
    \centering
    \includegraphics[width=0.8\textwidth]{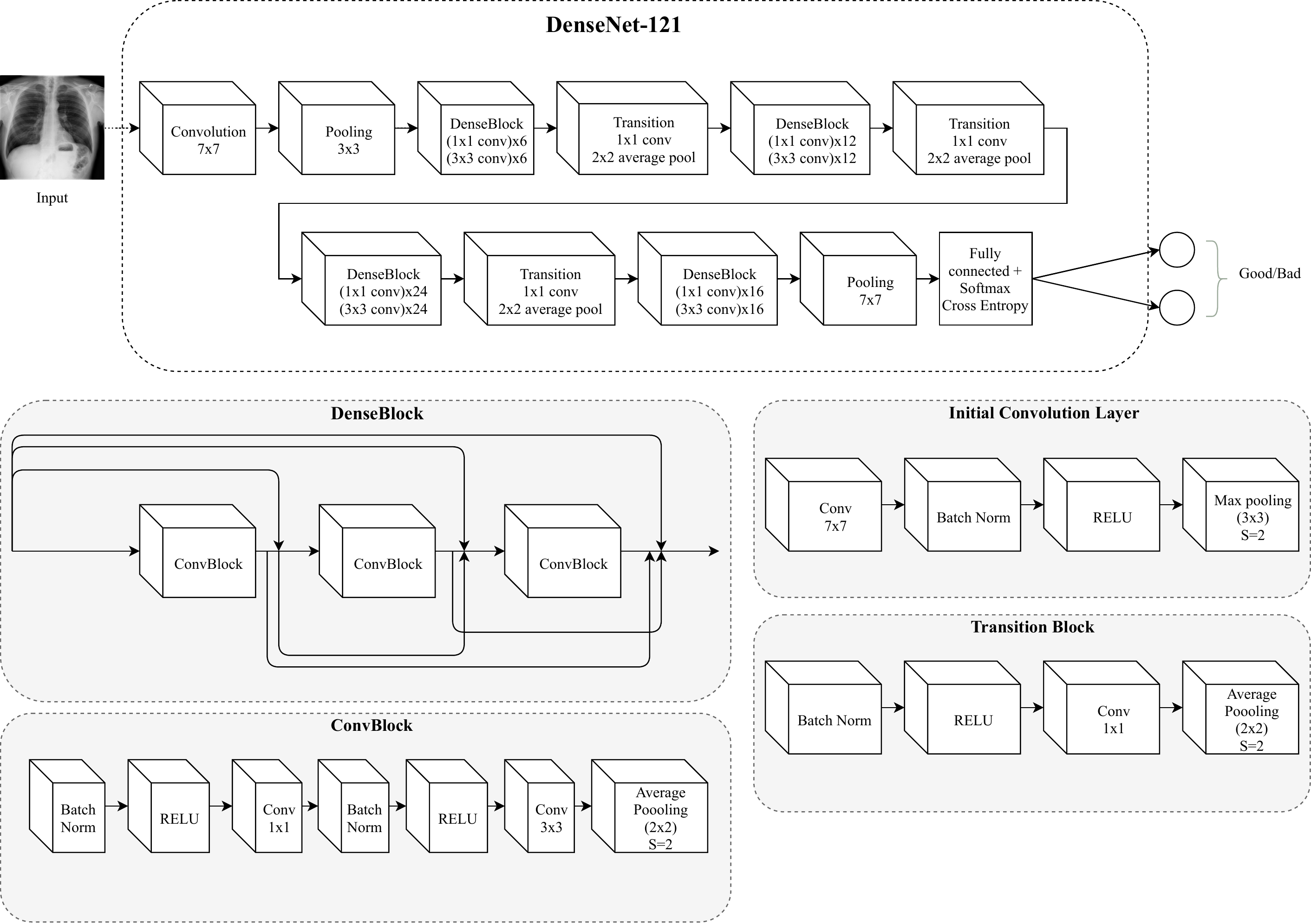}
    \caption{BXF architecture}
    \label{fig:bxf_architecture}
\end{figure*}
The BXF model is build with a 121-layer DenseNet like CheXNet but the softmax cross entropy is used as a loss function instead of the binary cross. Figure \ref{fig:bxf_architecture} shows the specific architecture of BXF model. The model contains 4 DenseBlocks and 3 TransitionBlocks situated in interleave positions. The initial Convolution layer has a kernel of 7x7. This layer includes Batch Normalization, RELU and 3x3 Max Pooling with stride of 2. Each DenseBlock has several densely connected convolution block, each output of convolution block is the input of all forward blocks. In every ConvolutionBlock, there are 1 Batch Normalization, 1 RELU, 1 1x1 Convolution, 1 Batch Normalization, 1 RELU, 1 3x3 Convolution and 1 2x2 Average Pooling with stride of 2. After each DenseBlock, a TransitionBlock is located with layers of Batch Normalization, RELU, 1x1 Convolution and 2x2 Average Pooling with stride = 2. The final fully-connected layer with softmax cross entropy give the 2-element vector output.

On CheXNet, the label is independent; the first-dimension sum of a sigmoid vector is not equal to 1, so the binary cross entropy is used. However, in this task, we classify images with only 2 labels, and the total probability of those labels must be 1.0, so the softmax function can work in this case. 

Softmax cross entropy function:
\begin{flalign}
&L(W;X,Y) = -\sum_{i = 1}^N \sum_{j = 1}^C y_{ji}\log(\hat{y}_{ji}) \\ 
&= -\sum_{i = 1}^N \sum_{j = 1}^C y_{ji}\log\left(\frac{\exp(w_j^Tx_i)}{\sum_{k=1}^C \exp(w_k^Tx_i)}\right) \\
&= -\sum_{i = 1}^N \sum_{j = 1}^C y_{ji}\log\left(softmax(W, x_i)\right)
\end{flalign}
with \(W\) is training weights, \(X\) is input, \(Y\) is label, \(N\) is number of data, \(C\) is number of classes (2 in this case). 

%% file: experiment.tex
\subsection{Bone shadow exclusion}
Using 247 pairs of X-ray images at JSRT \cite{dataset:jsrt} and BSE-JSRT \cite{dataset:bsejsrt}, we get 4,199 pairs of images from augmentation with a combination of affine transformations, including horizontal flipping, random zooming, shearing, shifting and rotating. We have trained in 150 epochs, each contains 800 steps, the initial learning rate is 0.001 and decay 25\% after 100 epochs. Besides that, we chose a batch size of 5 and standard \(\alpha=0.84\). 

This model took about 24 hours of training on a machine which has 8 CPUs, 12GBs RAM, 1 GPU Tesla K80 12GB. The final loss value is approximately equal to the published research. Particularly, the MS-SSIM value is 0.1075 while the MSE is 0.138, combined loss is 0.0925.

Available in approximately 1000 bounding boxes, we evaluated heatmaps of 346 images from the BSE and the original version of ChestX-ray14 dataset. These images are cropped to bounding box areas, then calculating mean CAM values from these regions. There are 181/346 BSE images has higher value in bounding box region than original ones.

\begin{figure}[!htb]
	\begin{center}
    	\includegraphics[width=0.49\columnwidth]{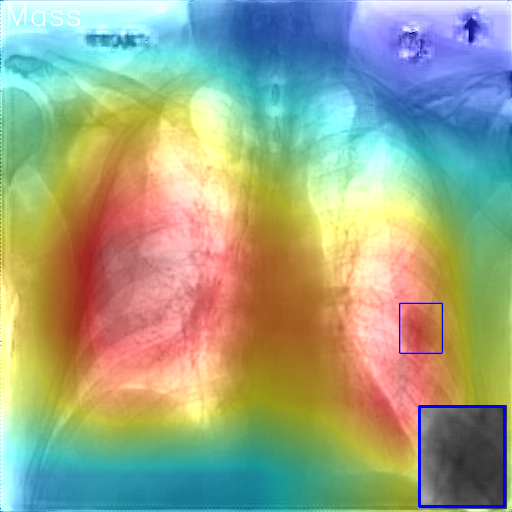}
        \includegraphics[width=0.49\columnwidth]{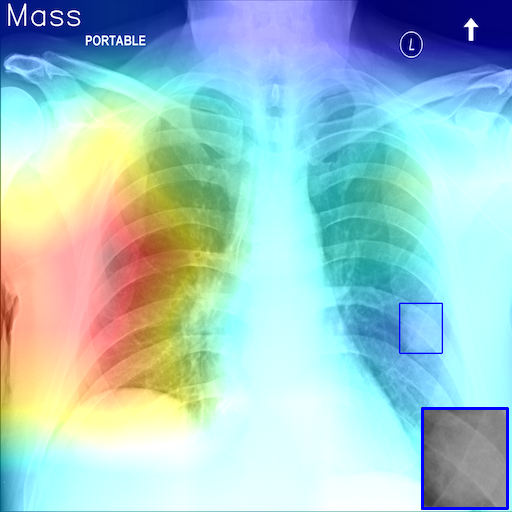}
        \caption{Heatmap of bone suppression (left) and origin (right) images}
        
        \label{heatmap_example}
	\end{center}
\end{figure}

Figure \ref{heatmap_example} shows that the CheXNet model can detect injured regions in bone suppression better than the original images. Blue-border rectangles are the groundtruth provided by radiologists in ChestX-ray14, in BSE image, the rectangle contains more red color than in the original one. Bottom right crops indicate that bone shadows in this region hide underneath injuries.

\begin{figure}[!htb]
	\begin{center}
    	\includegraphics[width=0.32\columnwidth]{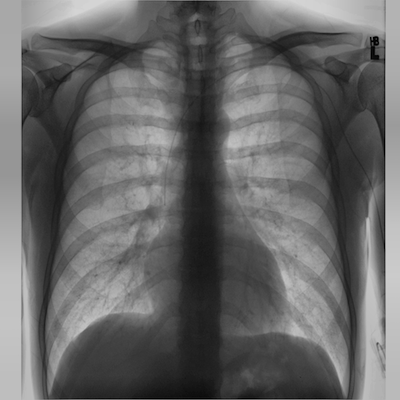}
        \includegraphics[width=0.32\columnwidth]{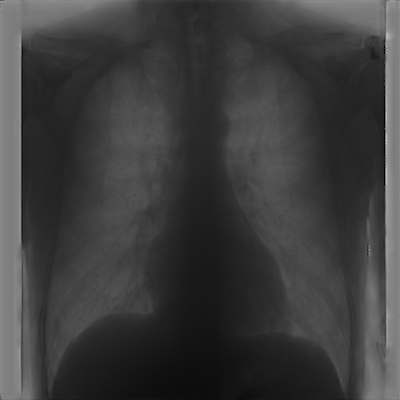}
        \includegraphics[width=0.32\columnwidth]{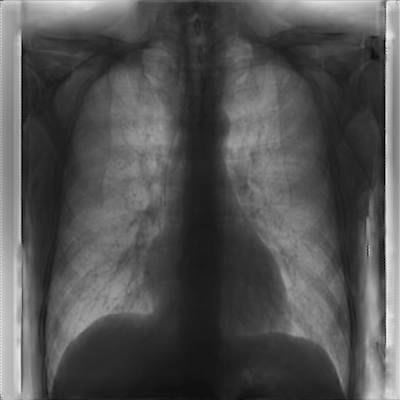}
        \caption{From left to right: Origin X-ray with brightness and color enhancement; BSE image; BSE-CLAHE image}
        \label{compare_bse_clahe}
	\end{center}
\end{figure}

{\renewcommand{\arraystretch}{1.2}

\begin{table*}[!htb]
	\caption{Comparing AUROC of different models}
      \centering
      \begin{tabu}to 0.95\textwidth { X[4,l] X[3.5,c] X[3.5,c] X[3.5,c] X[3.5,c] X[3.5,c] X[3.5,c] }
        \hline
        Pathology & CheXNet(2017)  & Our Implement & BSE & BSE-CLAHE & BXF\\ 
        \hline \hline
        Atelectasis & 0.8094 & 0.8250 & 0.8063 & 0.8107 & 0.8322 \\
        Cardiomegaly & 0.9248 & 0.9079 & 0.9062 & 0.9065 & 0.9114 \\
        Effusion & 0.8638 & 0.8846 & 0.8750 & 0.8749 & 0.8883 \\
        Infiltration & 0.7345 & 0.7142 & 0.6961 & 0.7000 & 0.7243 \\
        Mass & 0.8676 & 0.8592 &  0.8349 & 0.8229 & 0.8566 \\
        Nodule & 0.7802 & 0.7872 & 0.7672 & 0.7646 & 0.7891 \\
        Pneumonia & 0.7680 & 0.7590 & 0.7466 & 0.7605 & 0.7573 \\
        Pneumothorax & 0.8887 & 0.8832 & 0.8705 & 0.8775 & 0.8877 \\
        Consolidation & 0.7901 & 0.8170 & 0.8080 & 0.8015 & 0.8170 \\
        Edema & 0.8878 & 0.8932 & 0.8971 & 0.8959 & 0.8995 \\
        Emphysema & 0.9371 & 0.9280 & 0.9013 & 0.8972 & 0.9210 \\
        Fibrosis & 0.8047 & 0.8410 & 0.8243 & 0.8313 & 0.8417 \\
        Pleural Thickening & 0.8062 & 0.7908 & 0.7804 & 0.7698 & 0.7867 \\
        Hernia & 0.9164 & 0.9161 & 0.9254 & 0.9305 & 0.9103 \\
        \hline
        \textbf {Average} & \textbf{0.8414} & \textbf{0.8438} & \textbf{0.8301} & \textbf{0.8325} & \textbf{0.8445}\\
        \hline
      \end{tabu}
    \label{bse_clahe_evaluation}
\end{table*}}

\subsection{Contrast enhancement}

With contrast limitation is 3.0 and window size of 8x8, we use CLAHE for contrast enhancement on all BSE images. Figure \ref{compare_bse_clahe} shows different versions of a chest X-ray image. Contrast enhancement improved the quality of BSE image by highlighting the lung region without bone shadow like the original image.

To evaluate the performance of this task, we trained the BSE and BSE-CLAHE images with CheXNet models, got the result in Table \ref{bse_clahe_evaluation}. There are 7 of 14 diseases that have better AUROC after contrast enhancement, boosting the average AUROC by 0.0024. 

\subsection{CheXNet}

We trained two models with different versions of dataset: the original and the BSE. There are three subsets for each version: train (78,468 images), validation (11,219 images), and test (22,433 images). These sets are randomly splitting and be ensured that there is no patient overlap between the splits.

We used a machine with 16 CPUs, 60 GBs RAM and 1 GPU Tesla P100. These trained models are used for choosing strategies in the whole workflow and the results of them are shown in Table \ref{bse_clahe_evaluation}. 

Parameters for every model are set following the published research. The weights of ImageNet are also used as initialization. Adam with standard \(\beta_1\), \(\beta_2\) optimizes the model with the initial learning rate of 0.001. This learning rate is decayed by the factor of 10 after 10 epochs not decreasing validation loss. We use a batch size of 16 and it takes 22 hours for 100 epochs.

\subsection{BXF}
To label the ChestX-ray14 dataset for training BXF model, we get the sigmoid value of two models: one is trained with original data, and the other is trained with preprocessing images. Sigmoid value is a 14-dimensional vector, whose each element is the probability of a disease. We use Euclidean distance function to calculate the difference between the trained models \(u\) and the groundtruth \(v\), which is produced by ChestX-ray14. The lower of this value, the better of the model:
\[||u-v||_2 = \sqrt[]{\sum_{i=1}^{14}(u_i-v_i)^2}\]
Using this formula on each individual image of each trained model, we choose the best strategy for each piece of input data based on the lower value. The lower value between two models is labeled as 1, and the other get 0 as a label.

Before putting the images into the network, we downscale them to 224x224 and normalize based on the mean and standard deviation of images in the ImageNet training set. The weights of the network are also initialized  with weights from a model pretrained on ImageNet. The network is trained end-to-end using Adam with standard parameter (\(\beta_1=0.9\) and \(\beta_2=0.999\)) \cite{kingma2014adam}. We train the model using mini-batches of size 16. We use an initial learning rate of 0.001 that is decayed by a factor of 10 each time the validation loss plateaus after an epoch, and pick the model with the lowest validation loss. 
Using the train, validation and test splits in Table \ref{origin_bse_result_compare}, the machine with 8 CPUs, 12 GBs RAM, and 1 GPU Tesla K80 was executed. We also configure the Adam with standard parameters and initial learning rate is 0.0001 and is decreased by 30\% after 2 unimproved-loss-value epochs. We got the accuracy of 84.2 \% and F1-score of 78.9\% with the examined threshold of 0.46 (value \(>\) 0.46 will be labeled as 1) in test set after 30 training epochs.

\balance
In Table \ref{bse_clahe_evaluation}, the BXF model has better result than the CheXNet's AUROC published. There are 6 of 14 diseases that have better AUROC value than previous work.
{\renewcommand{\arraystretch}{1.2}
\begin{table}[!htb]
	\centering
    \caption{The evaluation results in 3 subsets. (1): Get better result with preprocessing; (2): Get better result without preprocessing; (3): Proportion of (2) in the whole subset}
      \begin{tabu}to 0.95\columnwidth { X[c] X[c] X[c] X[c]}
        \hline
        Result & Train & Validation & Test \\ 
        \hline \hline
        (1) & 31,310 & 4,718 & 8,685 \\
        (2) & 47,158 & 6,501 & 13,748\\
        (3) & 60.1\%  & 57.9\% & 61.3\% \\
        \hline
      \end{tabu}
    
    \label{origin_bse_result_compare}
\end{table}}

%% file: conclusion.tex
Lung diseases account for a significant of patient of morbidity and mortality. Early diagnosis normal lung diseases and treatment is very important to prevent complications, especially cause of cancer. Chest X-rays are the most effective examination used in diagnosis because of its cost and time. However, only one third of the global population can access radiology diagnostics \cite{mollura2010white}. Besides that, to interpret X-ray images, experts are required with several years of experience. With the development of deep neural networks, computers can easily support doctors in imaging tests, and CheXNet is one of those.

We develop additional steps to CheXNet for improving the current accuracy of that model. Bone shadow is excluded from the original X-ray images with contrast enhancement. However, this strategy does not always work well because of the lack of data in the bone suppression model. We also build a model to predict the accuracy of a raw image whether the preprocessing is applied. Our flow helps improve the heatmap of disease localization as well as increase average AUROC from 0.8414 to 0.8445.

However, there are existed points that can improve in our system. In the following works, we will train 2 CheXNet models with new split dataset: one contains only BSE images that really have good result and one includes original images which have high accuracy with released CheXNet model. However, we can improve the bone suppression model by training more data related to ChestX-ray14.

%% file: ack.tex
We would like to acknowledge Andrey G. for training script support.

This research is partially supported by the research funding from Honors Program, University of Science, Vietnam National University - Ho Chi Minh City.